\documentclass[fleqn,12pt,twoside]{article}
\usepackage{espcrc1}
\usepackage{epsfig}
\usepackage{times}
\newcommand{\AmS}{{\protect\the\textfont2
  A\kern-.1667em\lower.5ex\hbox{M}\kern-.125emS}}

\title{Elementary Hadronic Interactions at the CERN SPS
}

\author{H.G.~Fischer
\address[MCSD]{CERN, Geneva} for the NA49 Collaboration%
}

\begin{document}

\maketitle


\section{Introduction 
}
An extensive set of new data concerning elementary hadron+hadron 
and hadron+nucleus interactions has been accumulated over the past
few years at the CERN SPS at $\sqrt{s}=17.2$ GeV, using the NA49 detector.
These data, summarized in Table~1, contain more than 9 million
events with wide acceptance coverage and particle identification using
energy loss measurements in a Time Projection Chamber tracking system
\cite{nim}. 

\begin{table}[h]
\renewcommand{\tabcolsep}{2pc} 
\renewcommand{\arraystretch}{1.2} 
\vspace{-5mm}
\begin{center}
\begin{tabular}{|c|c|}
\hline
Reaction     & Events [M]  \\
\hline
p + p        & 5.000  \\    
n + p        & 0.120  \\
\hline
$\pi^+$ + p  & 0.640  \\
$\pi^-$ + p  & 0.450  \\
\hline
p + Pb       & 2.200  \\    
\hline
$\pi^+$ + Pb & 0.500  \\
$\pi^-$ + Pb & 0.480  \\
\hline
\end{tabular}
\caption{}
\end{center}
\vspace{-15mm}
\end{table}

A few general features of these data appear noteworthy:

\begin{itemize}
\item
\vspace{-3mm}
All hadron+nucleus reactions have centrality control via detection
of grey protons.
\item
\vspace{-3mm}
Complementary data samples with non-baryonic projectiles,
e.g. $\pi$-beams, allow for
the separation of target contributions as far as net baryon number
is concerned; in particular these data are obtained with opposite
projectile charge such that proper isospin averages may be performed.
\item
\vspace{-3mm}
A small but clean sample of n+p interactions with inverse kinematics
(neutron projectile) permits a first study of neutron fragmentation into
identified charged hadrons.
\end{itemize}

This new effort in a field that had been practically abandoned since many
years, is aimed at establishing an improved experimental basis for the study of
the evolution from elementary to nuclear hadronic interactions, in a model
independent fashion. For this aim, good phase-space coverage and particle
identification, completeness and complementarity with respect to reaction
channels are necessary pre-requisites. This is especially valid for a
renewed scrutiny of hadron+nucleus interactions which constitute the only
laboratory for elementary multiple collision processes.

In this short review, evidently only a very limited range of subjects can
be touched upon. Longitudinal and transverse momentum distributions of
baryons, the effects of isospin symmetry in neutron fragmentation, and
the yields of strange mesons and baryons will be discussed. Whenever
relevant, results from Pb+Pb collisions obtained with the same detector
will be shown for comparison.

\section{The Evolution of Longitudinal Baryon Density and Hadronic Factorization
}
The longitudinal transfer of baryon number in hadronic collisions from
target and projectile momenta to the central region constitutes one of
the most important observables as it has to do with final state energy
density ("stopping"). A central argument in this context will be the
factorization between target and projectile components, i.e. the way how
these contributions superimpose, how they may be experimentally separated 
and put together again in the more complex, asymmetric case of 
hadron+nucleus collisions.

To this intent in Fig.~1a the net proton density as a function of Feynman $x_F$ for
proton+proton collisions is tentatively separated
into a target and a projectile component meeting by
symmetry at half-height at $x_F=0$. 
Using now isospin-averaged pion
projectiles the net proton density as a function of $x_F$ 
can be obtained for a  net baryon number zero projectile (Fig.~1b). 
The ratio of the
pion-induced to proton-induced densities (Fig.~1c) should be 1 in the target hemisphere
and pass through 0.5 at $x_F=0$ if factorization, and herewith independence of
baryon transfer on projectile type, holds.

\begin{figure}[h]
\vspace{-6mm}
\centering
\includegraphics[width=14cm]{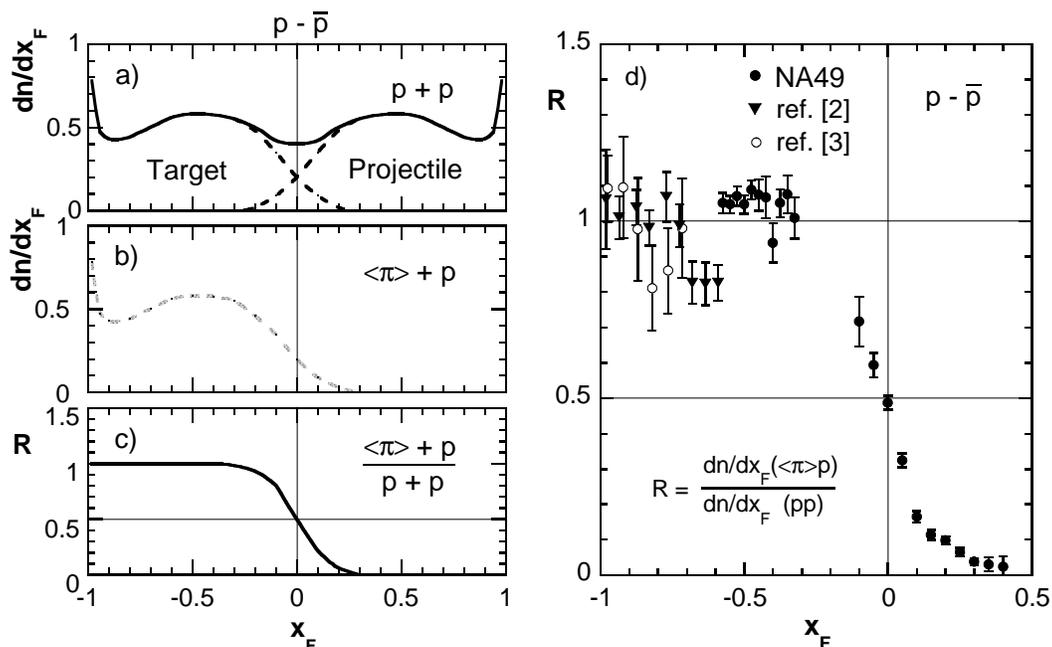}
\vspace{-14mm}
\caption{Schematic view of a) factorization of the net proton density in target and 
projectile components, b) net proton density for a  net baryon number zero projectile,
c) ratio of the pion-induced to proton-induced densities, and d) data on this ratio
(NA49, \cite{aji}, \cite{whi}).  
}
\label{f1}
\end{figure}

\vspace{-8mm}
The data shown in Fig.~1d clearly support this picture:

\begin{itemize}
\item
\vspace{-2mm}
The density ratio is 1 in the target hemisphere, i.e. for $x_F \leq -0.3$.
\item
\vspace{-3mm}
The ratio passes through 0.5 at $x_F=0$. 
It should be stressed that this is a non-trivial result.
\item
\vspace{-3mm}
The rapid approach to zero in the projectile hemisphere  -- together with
the integral of the distribution -- does not leave room for an important
"leakage" of baryon number from the target to the projectile side. 
This puts constraints e.g. on the applicability of the so-called "junction mechanism".
\end{itemize}

This argumentation may now be extended to nuclear data by using 
isospin-averaged pion+nucleus interactions at different centralities. As shown
in Fig.~2a, the net proton density in the forward region -- which is not
affected by nuclear rescattering at SPS energy -- shows a smooth and
conformal increase with the mean number of collisions $\nu$ suffered by the
projectile. This is direct experimental evidence for target pile-up which is
demonstrated to be linear with $\nu$ at $x_F=0$ in Fig.~2b.

\begin{figure}[h]
\vspace{-5mm}
\centering
\includegraphics[width=16cm]{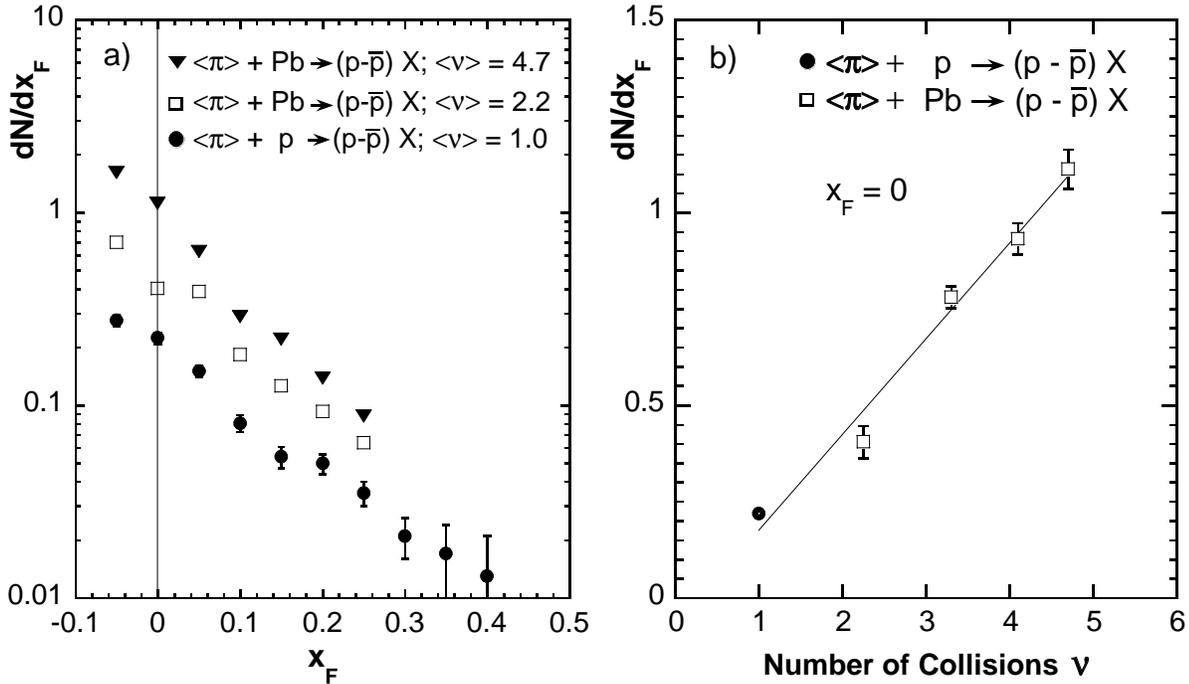}
\vspace{-12mm}
\caption{a) Net proton density as function of $x_F$ for various mean numbers 
of collisions $\nu$; b) net proton density at $x_F=0$ as function of $\nu$.
}
\label{f2}
\end{figure}

\vspace{-8mm}
This result opens up the possibility to obtain experimentally the projectile
contribution to the net baryon density in p+Pb interactions and thereby 
a clear picture of baryon number transfer in multiple
hadronic collisions. In fact by subtracting the target component (Fig.~2a)
from the total net baryon density in p+Pb collisions (Fig.~3a) for
the same number of projectile collisions, the projectile component
shown in Fig.~3b can be extracted (see also \cite{ryb}). 

The strong increase of this projectile contribution for p+Pb collisions 
with respect to the 
elementary p+p reaction is quantified in Fig.~4 as function of $\nu$.
 
\begin{figure}[t]
\centering
\includegraphics[width=14cm]{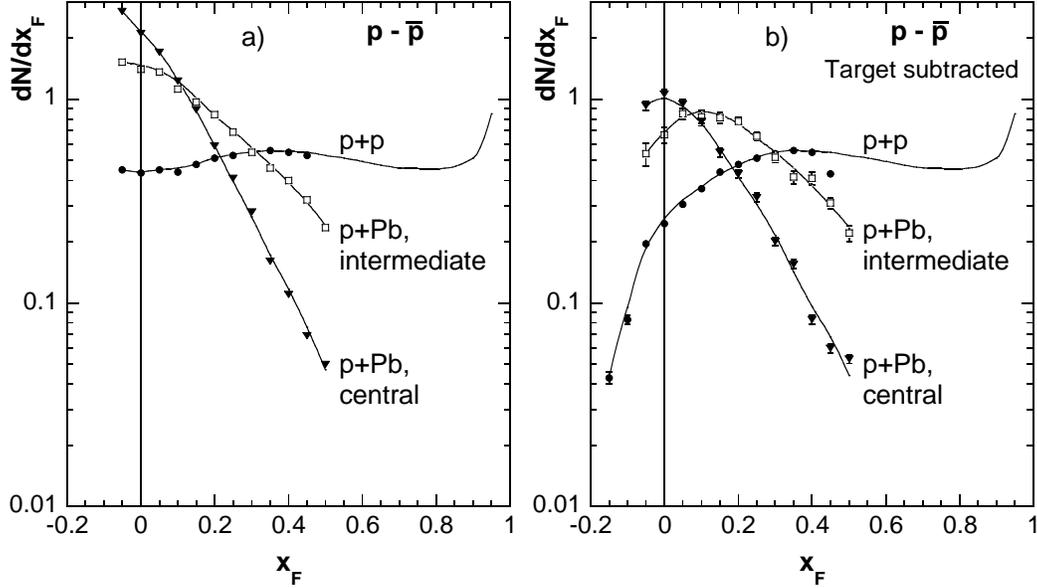}
\vspace{-11mm}
\caption{a) total net baryon density as function of $x_F$; b) projectile component
of net baryon density.
}
\label{f3}
\vspace{-8mm}
\end{figure}

Similar measurements should be performed for all other types of "net baryons",
e.g. $\Lambda, \Xi, \Omega$, in order not to confound an increase from baryon
number transfer with true nuclear enhancement factors (see chapter 4 and the
contribution from M.Kreps \cite{kre} to these proceedings).

As far as the (symmetric) Pb+Pb interactions are concerned, their central net
proton density may be directly compared to the (symmetric) p+p collisions
in order to extract the effect of baryon stopping. This is also shown for
peripheral and central Pb+Pb interactions in Fig.~4 (see \cite{ryb} for a more
detailed argumentation). Two important results emerge:

\begin{itemize}
\item
\vspace{-3mm}
There is a smooth increase of central net proton density with the number
of projectile collisions.
\item
\vspace{-3mm}
Taking into account the uncertainty in determining $\nu$ 
there is no difference between p+Pb and Pb+Pb interactions.
\end{itemize}

\begin{figure}[h]
\vspace{-11mm}
\centering
\includegraphics[width=14cm]{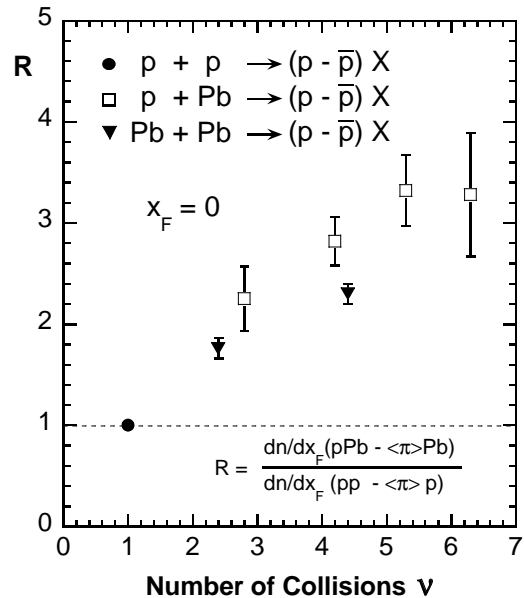}
\vspace{-11mm}
\caption{Projectile contribution to the net proton density at $x_F=0$ as function of
$\nu$ for p+Pb collisions normalized to p+p interactions.
}
\label{f4}
\end{figure}

\section{The Evolution of Transverse Baryon Density
}
Given the substantial modification of longitudinal baryon distributions in 
multiple hadronic collisions as discussed above, it is interesting to regard
also the transverse baryonic density distributions. In fact it would be rather
surprising if such strong long-range longitudinal effects did not leave
their trace also in the transverse dimension.
  
A first indication may be obtained by comparing the transverse proton density
distributions $dn/dx_Fdp_T$ at $x_F=0$ for p+p and central p+Pb interactions as
shown in Fig.~5a. A considerable shift to higher transverse momenta is evident
in p+Pb collisions. This corresponds to the well-known Cronin effect \cite{cro},
here however measured for about 5 rather than 3.7 average projectile
collisions. Following the argumentation of the preceding chapter we may now
extract the relative contributions of target and projectile to this
effect.

\begin{figure}[h]
\vspace{-8mm}
\centering
\includegraphics[width=15.5cm]{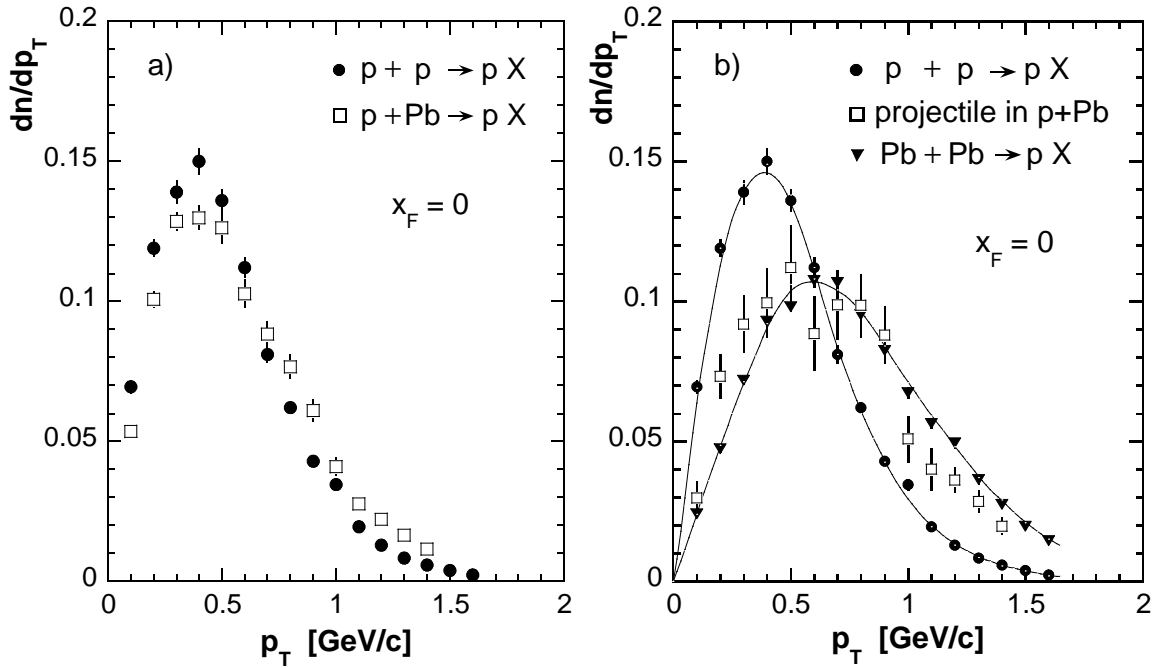}
\vspace{-10mm}
\caption{a) $p_T$ distribution of protons at $x_F=0$ for p+p and p+Pb interactions;
b) $p_T$ distribution of protons at $x_F=0$ for p+p and projectile component to this 
distribution for p+Pb compared to $p_T$ distribution for Pb+Pb interactions.
}
\label{f5}
\end{figure}

\vspace{-6mm}

The target contribution should be represented by the transverse net proton
distribution from $\langle\pi\rangle$+Pb collisions which turns out to be equal to
the distribution from p+p interactions. This leaves the projectile
contribution as the source of the observed $p_T$ broadening and indeed a further 
substantial upward shift in transverse momentum is observed as seen in Fig.~5b
(open squares). This projectile contribution approaches the proton $p_T$ 
distribution obtained in central Pb+Pb collisions also shown in Fig.~5b.  

Several conclusions may be drawn from this study:

\begin{itemize}
\item
\vspace{-3mm}
Again the results from central p+Pb and Pb+Pb collisions are very
similar.
\item
\vspace{-3mm}
A new interpretation of the Cronin effect as a two-component phenomenon
including a transversally inert target contribution  
is necessary at least in the $p_T$ range considered here.
\item
\vspace{-3mm}
Major transverse activity is induced by multiple hadronic collisions
already on the level of elementary p+A interactions.
\end{itemize}

The last conclusion casts some doubt on the validity of the "standard"
interpretation of transverse momentum broadening in heavy ion collisions
by transverse expansion of a "hot" initial partonic phase. 
Doubtlessly by introducing a vector field of expansion of sufficient
complexity it will always be possible to map two different
transverse momentum distributions onto each other. Such a vector field has
however no place in p+A interactions. It should rather be asked what happens to
transverse hadronization in multiple collision processes as these processes
establish the link between p+A and A+A collisions.

Finally it should be mentioned here that the hierarchy of transverse
momentum broadening with respect to particle mass, with pions showing
practically no effect up to $p_T \simeq 1$~GeV/c, 
is also reproduced in p+A collisions.

\section{Consequences of Isospin Symmetry on Kaon and Baryon Production
}
In view of the fact that neutrons constitute 60\% of the baryonic participants
in heavy ion interactions, the almost complete absence of experimental
information on neutron fragmentation constitutes a major flaw. Although
isospin invariance is of course uncontested in hadronic interactions,
its consequences for final state hadronization depend on details of production
mechanisms and are as such incalculable; they have to be extracted experimentally.

A first measurement of charged pion and kaon yields from neutrons in the
projectile hemisphere, see Fig.~6, may illustrate these problems.

\begin{figure}[h]
\vspace{-5mm}
\centering
\includegraphics[width=16cm]{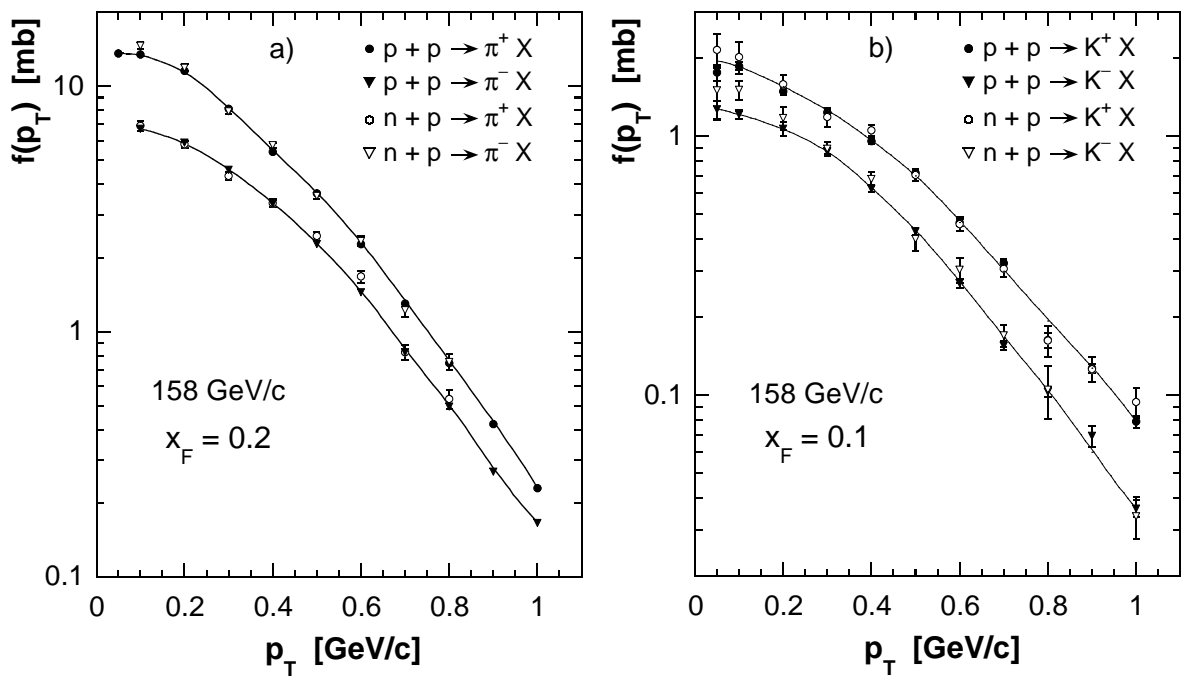}
\vspace{-10mm}
\caption{Differential invariant cross section as function of $p_T$ for 
a) $\pi^+$ and $\pi^-$ 
b) K$^+$ and K$^-$ produced in p+p and n+p interactions. 
}
\label{f6}
\end{figure}

\vspace{-5mm}
Apparently $\pi^+$ and $\pi^-$ change their place when switching from p to n
projectile, but K$^+$ and K$^-$ do not (within error bars). Whereas the first
fact may be expected from the relevant isospin multiplets given in Table~2,
the latter one can only be understood by the prevalence of associate
K-hyperon production at SPS energy, and even so only by invoking important
high-mass N$^*$, Y$^*$ and K$^*$ formation and subsequent decay.

\begin{table}[h]
\vspace{-7mm}
\begin{center}
{\small
\begin{tabular}{|l|c|c|c|c|c|}
\hline
\multicolumn{6}{|l|}{Isospin $I=1$}                                 \\
\hline
   Projectiles             &           &  n  &           &  p  &             \\
\hline
   Produced particles      &  $\pi^-$  &     &  $\pi^0$  &     &  $\pi^+$    \\    
\hline
   $I_{\mbox{3}}$          &   -1    & -1/2 &   0   & 1/2 &   +1       \\
\hline
\end{tabular}

\begin{tabular}{|l|c|c|c|}
\hline
\multicolumn{3}{|l|}{Isospin $I=1/2$}                             & Strangeness  \\
\hline
   Projectiles          &   n     &    p                      &             \\
\hline
   Produced particles   &  K$^0$  &  K$^+$                    &  1          \\    
   (associate)          &  K$^-$  &  $\overline{\mbox{K}}^0$  &  -1         \\
\hline
   $I_{\mbox{3}}$       &  -1/2   &  +1/2                     &             \\
\hline
\end{tabular}
\begin{tabular}{|l|c|c|c|c|c|c|}
\hline
\multicolumn{6}{|l|}{Isospin $I=1$}                             & Strangeness  \\
\hline
  Projectiles         &           & n &          & p &          &             \\
\hline
  Produced particles  & K$^-$K$^0$ & & K$^+$K$^-$ & & K$^+$$\overline{\mbox{K}}^0$ & 0 \\  
  (in pairs)          &      & & K$^0$$\overline{\mbox{K}}^0$ & &   &                   \\
\hline
  $I_{\mbox{3}}$      & -1 & -1/2 &          0              & 1/2 & +1 &   \\
\hline
\end{tabular}
}
\end{center}
\vspace{-3mm}
\caption{}
\vspace{-14mm}
\end{table}

This experimental observation has strong consequences for the interpretation
of K/$\pi$ ratios when comparing elementary to nuclear collisions. The necessary
isospin correction factors for a 60/40\% n/p mixture with respect to p+p
interactions are presented in Fig.~7a, the corresponding isospin-corrected
K$^+$ enhancement factors in Fig.~7b for central p+Pb and Pb+Pb collisions,
as a function of $x_F$. Again one finds, together with a smooth dependence on the 
number of collisions (not shown here), no apparent difference between p+A and A+A 
interaction. A similar result applies to K$^-$ production.

\begin{figure}[h]
\vspace{-8mm}
\centering
\includegraphics[width=14cm]{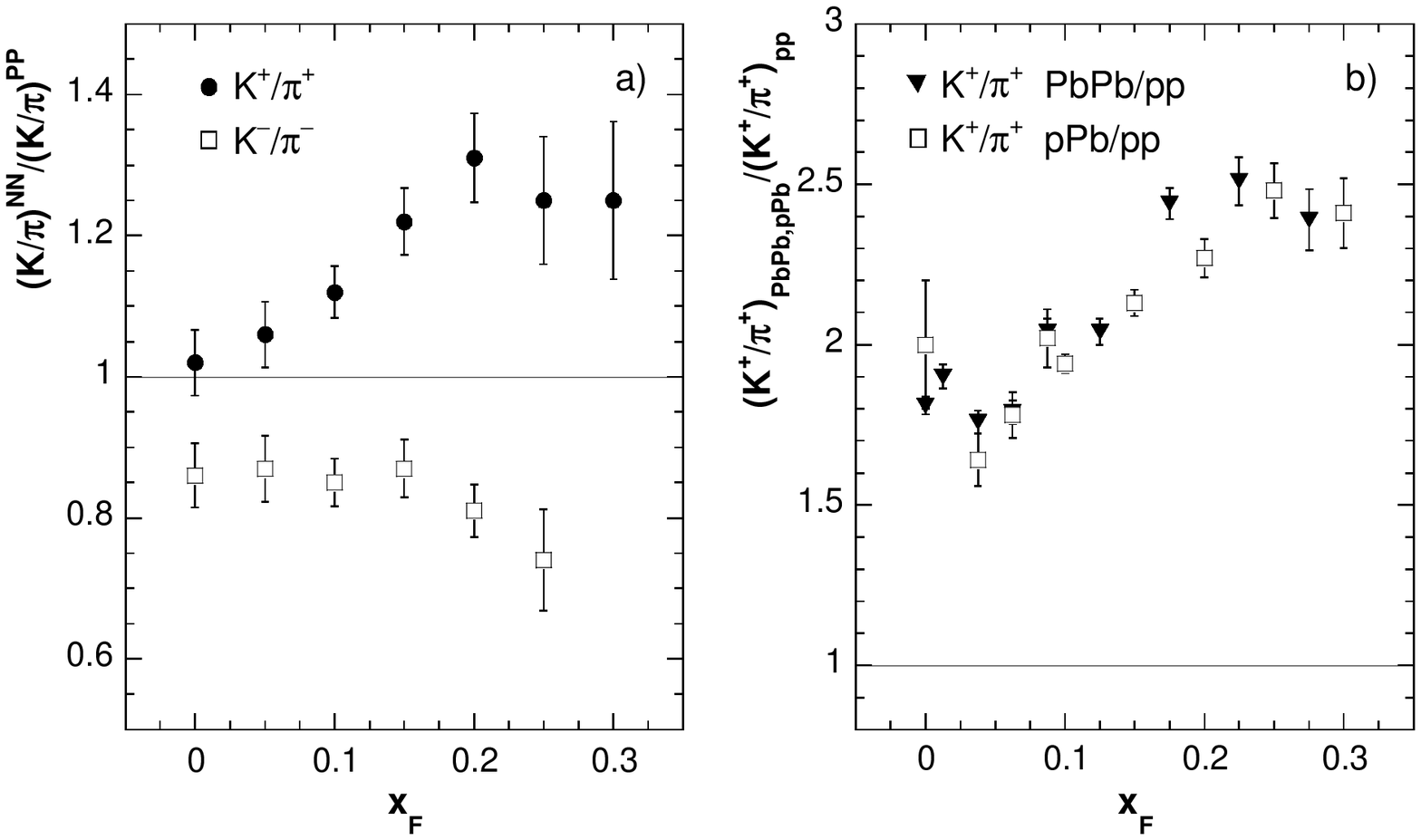}
\vspace{-10mm}
\caption{a) K/$\pi$ ratio as function of $x_F$ for N+N collisions, 
N standing for a 60/40\% neutron/proton mixture, normalized to p+p interactions; 
b) isospin corrected K$^+$/$\pi^+$ ratio as function of $x_F$ for p+Pb and Pb+Pb 
collisions, each normalized to p+p interactions.
}
\label{f7}
\end{figure}


Concerning baryon pair production, the anti-proton yield from n+p as compared 
to p+p collisions, shown in Fig.~8a, constitutes yet another non-trivial 
consequence of isospin symmetry. 
In fact there is a sizeable increase of anti-proton
production from neutrons which amounts to about a factor of 1.5 
if taking proper account of the equal target in both reactions. 


This increase would be difficult to hide in any (symmetric) sea contribution
to baryon pair production. Instead one has to allow for an important yield
of asymmetric baryon pairs of the type p$\overline{\mbox{n}}$ or 
$\overline{\mbox{p}}$n following from the the
corresponding isospin triplet, shown in Table~3.

\begin{table}[htb]
\renewcommand{\tabcolsep}{1.2pc} 
\vspace{-5mm}
\begin{center}
{\small
\begin{tabular}{|l|c|c|c|c|c|}
\hline
\multicolumn{6}{|l|}{Isospin $I=1$}                                 \\
\hline
    Projectiles             &         &   n  &       &  p  &         \\
\hline
    Produced particles  & $\overline{\mbox{p}}$n & & p$\overline{\mbox{p}}$ & & p$\overline{\mbox{n}}$       \\    
                           &         &      & n$\overline{\mbox{n}}$ &   &        \\
\hline
   $I_{\mbox{3}}$          &   -1    & -1/2 &   0   & 1/2 &   +1       \\
\hline
\end{tabular}
}
\end{center}
\vspace{-3mm}
\caption{}
\vspace{-13mm}
\end{table}

This indicates production via
high mass mesonic states which seems the only way to understand the very
large correlation with projectile isospin observed, as such states might be
generated in a more primordial hadronization phase.

Asymmetric baryon pair production has immediate consequences for the definition
of "net" baryon yields. Evidently the usual difference p--$\overline{\mbox{p}}$ 
is not the correct measure: 
In p-fragmentation the $\overline{\mbox{p}}$ yield under-estimates the number
of pair produced protons, in n-fragmentation it over-estimates it. 
Indeed the usual p--$\overline{\mbox{p}}$ 
difference in the invariant, $p_T$ integrated cross section at $x_F=0$,
shown in Fig.~8b as function of $\sqrt{s}$, flattens out in the ISR energy range.
This contradicts baryon number conservation. Taking account of the
proper number of pair produced protons by subtracting 1.6 times the
$\overline{\mbox{p}}$
yield, also shown in Fig.~8b, this contradiction is lifted and the "net"
baryon yield approaches zero at the highest ISR energies.

\begin{figure}[h]
\vspace{-8mm}
\centering
\includegraphics[width=14cm]{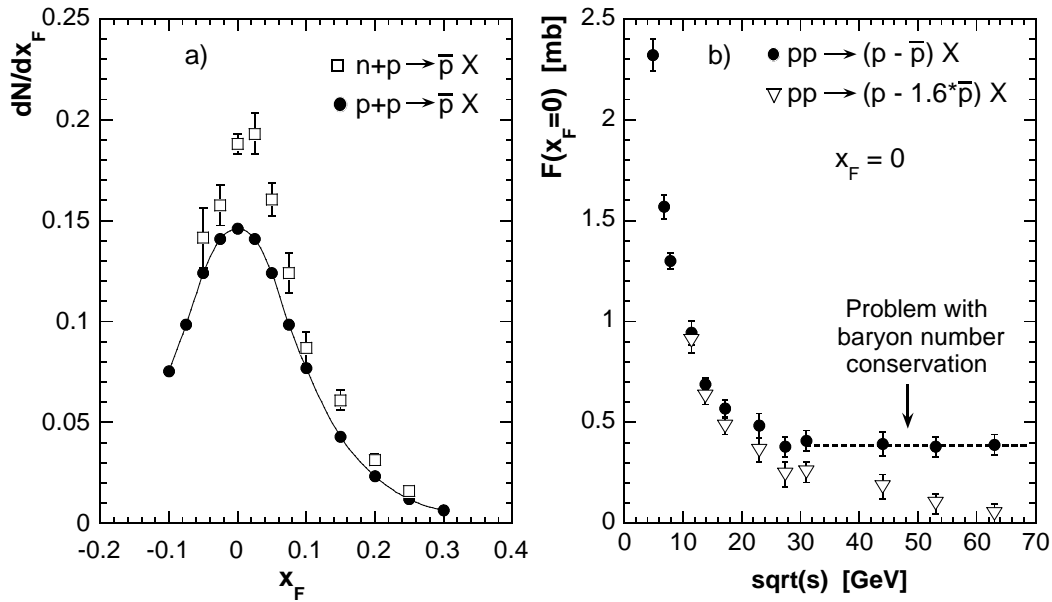}
\vspace{-10mm}
\caption{a) Anti-proton density as function of $x_F$ for n+p and p+p interactions; 
b) invariant cross section at $x_F=0$ as function of $\sqrt{s}$ for net protons produced
in p+p interactions.
}
\label{f8}
\end{figure}

\section{A Comment on Strange and Multistrange Baryon Poduction
}
In the realm of "signatures" for "new" physics in heavy ion collisions,
strange and multistrange baryons occupy a special position: 
Enhancement factors of about 
3 for $\Lambda$, 10 for $\Xi^-$ and 20 for $\Omega$ with respect to
"elementary" interactions have been reported. In view of the above discussion,
these evidences have to be scrutinized very carefully for different
systematic effects:

\begin{itemize}
\item
\vspace{-3mm}
It is a necessity to measure reference cross sections for elementary
baryon+baryon interactions: p+Be e.g. is in this sense not a reference.
\item
\vspace{-3mm}
It is a necessity to take proper account of isospin effects as shown above.
\item
\vspace{-3mm}
It is a necessity to quantify and subtract effects induced by baryon
number transfer. This needs wide acceptance coverage far beyond the
standard range close to central rapidity.
\item
\vspace{-3mm}
It is a necessity to study the elementary hadron+nucleus reaction as
this is the only way to get access to the effect of multiple collisions
in "cold" hadronic matter. This should be done with controlled centrality.
\item
\vspace{-3mm}
In this latter case it is a necessity to learn how to disentangle 
target and projectile contributions which are here highly asymmetric
compared to central nucleus+nucleus collisions.
\end{itemize}

In the case of cascade production, at least part of this experimental
program has been fulfilled by the NA49 experiment \cite{sus}. Cross sections for
p+p collisions have been measured as well as for p+Pb and Pb+Pb interactions.
Due to the premature closure of the SPS fixed target program it will not
be possible to work up the rest of the straight-forward experimental list
presented above. Here one has to take refuge to some (reasonable) assumptions.
A more detailed account of this approach is given in \cite{hgf1,hgf2} and
the contribution of M.~Kreps \cite{kre} to this conference.

The key to a proper understanding of nuclear enhancement of cascade baryons is
the last point quoted above. From the experience gained with target and
projectile contributions it seems more natural to blame a measured enhancement
in p+Pb on the projectile, which has undergone multiple collisions, rather
than on the target which has been shown to pile up with elementary cross
section at least for protons.

In doing so one arrives at the enhancement factors $E$ at
central rapidity (for definition see \cite{kre})
shown in Fig.~9 for
p+A collisions as function of the number of projectile collisions, including
data from WA97 \cite{w97} for minimum bias p+Be and p+Pb interactions. If compared
to the enhancement factors extracted from Pb+Pb reactions one sees again, within
reasonable error bars (especially on the scales concerning the number
of  collisions in both reactions), a very similar behaviour between the 
two types of nuclear collisions.

Another reasonable assumption may be made concerning isospin effects. In
writing down the isospin triplet of $\Xi\overline{\Xi}$ states, Table~4,
it appears that this triplet is charge-antisymmetric to the triplet of
$S=0$ baryons (see Table~3).

\begin{table}[h]
\vspace{-6mm}
\begin{center}
{\small
\begin{tabular}{|l|c|c|c|c|c|c|}
\hline
\multicolumn{6}{|l|}{Isospin $I=1$}                             & Strangeness  \\
\hline
   Projectiles             &         &   n  &       &  p  &     &                 \\
\hline
   Produced particles  & $\overline{\Xi}^0\Xi^-$ & & $\Xi^0\overline{\Xi}^0$ & & $\Xi^0\overline{\Xi}^+$ & -2, +2       \\    
                           &         &      &  $\Xi^-\overline{\Xi}^+$  &    &   &  \\
\hline
   $I_{\mbox{3}}$          &   -1    & -1/2 &   0   & 1/2 &   +1  &   \\
\hline
\end{tabular}
}
\caption{}
\end{center}
\vspace{-15mm}
\end{table}

\begin{figure}[t]
\centering
\includegraphics[width=16cm]{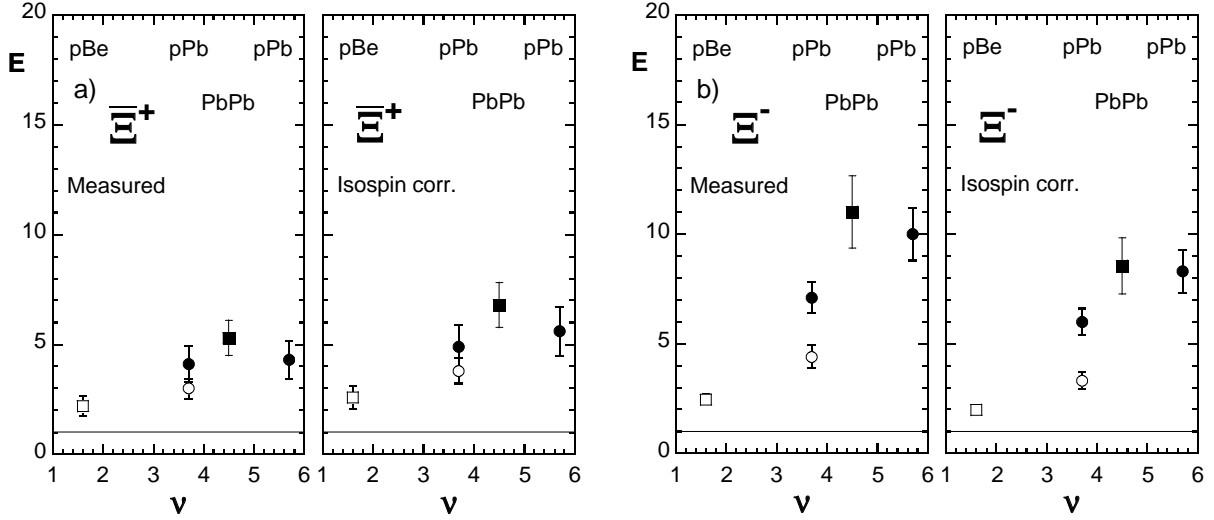}
\vspace{-12mm}
\caption{a) Enhancement factor $E$ at central rapidity for $\overline{\Xi}^+$ 
production as function
of $\nu$ in p+Be, p+Pb and 
Pb+Pb collisions normalized to p+p interactions and next to it the isospin corrected $E$
as decribed in text; 
b) same for $\Xi^-$.
}
\label{f9}
\end{figure}

\vspace{-15mm}
A correspondingly opposite dependence on projectile
isospin has therefore to be expected in the sense that
$\overline{\Xi}^+$ production 
should be enhanced in p-fragmentation as opposed to n-fragmentation. The
opposite trend should be valid for $\Xi^-$production. Assuming this asymmetry
to be of the same size as the one observed for $S=0$ baryons one arrives
at the isospin-corrected comparison of p+p, p+A and A+A interactions shown
in Fig.~9. Apparently the difference in enhancement of about a factor
of 2 observed between $\Xi^-$ and $\overline{\Xi}^+$ is now much 
reduced and the remaining
effect might well be attributed to stopping, see chapter~2 above \cite{kre}.

\section{Conclusion
}

Some features of an extensive set of new data in elementary hadron+nucleon
and hadron+nucleus interactions have been presented and compared to
nucleus+nucleus collisions. Except for the use of very fundamental
properties of hadronic reactions like baryon number conservation and
isospin invariance the main aim of this study is to provide a model
independent approach based exclusively on completeness and internal
consistency and complementarity of the data. Several conclusions may be drawn:

\begin{itemize}
\item
\vspace{-3mm}
Momentum distributions and yields of hadrons follow a smooth evolution from
elementary to nuclear collisions. In particular, nucleus+nucleus collisions
comply well with this evolution and do not occupy a special place.

\item
\vspace{-3mm}
Multiple hadronic collision processes are at the origin of this evolution;
the essential physics variable is the number of collisions per participant
or -- in a more general sense -- the depth of nuclear matter traversed by the
participant hadron.

\item
\vspace{-3mm}
Projectile isospin plays an important role; hence the necessity to study
especially neutron fragmentation in more detail.

\item
\vspace{-3mm}
Complete, high quality and high statistics data sets on elementary hadronic
interactions are indispensible in order to gain model independent
understanding of the more complex nuclear collisions. In the absence of
reliable theoretical approaches in the field of non-perturbative QCD
a better experimental basis has to be established. The present study
has hopefully demonstrated that such a basis may be obtained. In this sense
the detailed scrutiny of elementary hadronic collisions should be
vigorously pursued. 
\end{itemize}




\end{document}